\documentstyle[12pt]{article}
\input epsf
\begin{document}

\def\question#1{{{\marginpar{\small \sc #1}}}}
\newcommand{\QCD}{{ \rm QCD}^{\prime}}
\newcommand{\MSSM}{{ \rm MSSM}^{\prime}}
\newcommand{\eq}{\begin{equation}}
\newcommand{\en}{\end{equation}}
\newcommand{\bino}{\tilde{b}}
\newcommand{\tsquark}{\tilde{t}}
\newcommand{\gluino}{\tilde{g}}
\newcommand{\photino}{\tilde{\gamma}}
\newcommand{\wino}{\tilde{w}}
\newcommand{\mtilde}{\tilde{m}}
\newcommand{\higgsino}{\tilde{h}}
\newcommand{\gsi}{\,\raisebox{-0.13cm}{$\stackrel{\textstyle>}
{\textstyle\sim}$}\,}
\newcommand{\lsi}{\,\raisebox{-0.13cm}{$\stackrel{\textstyle<}
{\textstyle\sim}$}\,}

\rightline{RU-96-05}
\rightline{hep-ph/9603271}

\baselineskip=18pt
\vskip 0.7in
\begin{center}
{\bf \LARGE Detecting Gluino-containing Hadrons}\\
\vspace*{0.9in}
{\large Glennys R. Farrar}\footnote{Research supported in part by
NSF-PHY-94-23002} \\
\vspace{.1in}
{\it Department of Physics and Astronomy \\ Rutgers University,
Piscataway, NJ 08855, USA}\\
\end{center}
\vspace*{0.2in}
\vskip  0.9in  

{\bf Abstract:} When SUSY breaking produces only dimension-2
operators, gluino and photino masses are of order 1 GeV or less.  The
$g \tilde{g}$ bound state has mass 1.3-2.2 GeV and lifetime
${\,\raisebox{-0.13cm}{$\stackrel{\textstyle>} {\textstyle\sim}$}\,}
10^{-5} - 10^{-10}$ s.  This range of mass and lifetime is largely
unconstrained because missing energy and beam dump techniques are
ineffective.  With only small modifications, upcoming $K^0$ decay
experiments can study most of the interesting range.  The lightest
gluino-containing baryon ($uds \tilde{g}$) is long-lived or stable;
experiments to find it and the $uud \tilde{g}$ are also discussed. 
\thispagestyle{empty}
\newpage
\addtocounter{page}{-1}
\newpage

I have recently outlined\cite{f} some of the low-energy
features of theories in which dimension-3 SUSY breaking operators are
highly suppressed.  This is the generic situation in several
interesting methods of SUSY breaking.  Two to four free parameters of
the usual minimal supersymmetric standard model ($A$ and the gaugino
masses) vanish at tree level.  The elimination of these SUSY breaking 
operators implies that there is no additional CP violation at T=0 
beyond what is already present in the standard
model\cite{f}.  (In contrast, conventional SUSY-breaking
generically leads to the embarassing prediction of a neutron electric
dipole moment 3-4 orders of magnitude larger than the present
experimental upper limit.)  The allowed range of the remaining SUSY 
parameters can be constrained by requiring correct breaking of the
SU(2)$\times$U(1) gauge symmetry, consistency with LEP mass limits,
and the absence of any new flavor singlet pseudoscalar lighter than
the $\eta'$\cite{f}.  

Gauginos are massless at tree level but get calculable masses through
radiative corrections from electroweak (gaugino/higgsino-Higgs/gauge
boson) and top-stop loops. Evaluating these within the constrained
parameter space leads to a gluino mass range $m_{\tilde{g}}\sim
\frac{1}{10} - 1$ GeV and photino mass range $m_{\tilde{\gamma}} \sim
\frac{1}{10} - 1 \frac{1}{2}$ GeV.  The lightest chargino has a mass
less than $m_W$.  The photino is an attractive dark matter candidate,
with a correct abundance for parameters in the predicted 
ranges\cite{f:100}. Due to the non-negligible mass of the photino
compared to the glueball, prompt photinos are not a useful signature
for the light gluinos and the energy they carry\cite{f:95}.  Gluino
masses less than about $1 \frac{1}{2}$ GeV are largely
unconstrained\cite{f:95}. Experiments to rectify this are proposed
here.  Consequences for squark and chargino searches are discussed in
ref. \cite{f:105}.  

The gluino forms bound states with gluons and other gluinos, as well
as with quarks and antiquarks in a color octet state. The lightest of
these states, the spin-1/2 gluon-gluino bound state called $R^0$,
should have a mass $\sim 1.3 - 2.2$ GeV\cite{f:95,f}.  Since the
gluino is light, this state is approximately degenerate with a flavor
singlet pseudoscalar comprised mainly of $\tilde{g} \tilde{g}$\cite{f:95}.
Experimental evidence is now quite strong for an ``extra'' flavor
singlet pseudoscalar at $\sim 1500$ MeV\cite{etagluino}, in
addition to those which can be accomodated in ordinary
QCD\cite{f}.  The $\eta'$ is identified with the pseudogoldstone
boson associated with the breaking of the chiral $R$-symmetry of the
nearly massless gluino\cite{f:95}. The lightest $R$-baryon is the
flavor-singlet spin-0 $uds\tilde{g}$ bound state called $S^0$, whose
mass should lie $0-1$ GeV above that of the $R^0$. Higher lying
$R$-hadrons decay to the $R^0$ and $S^0$ via conventional strong or
weak interactions.  The rest of this paper is devoted to finding
evidence for these $R$-hadrons. 

I shall assume here that photinos are responsible for the cold dark
matter of the Universe.  This fixes more exactly the mass of the
photino and $R^0$ because in order to obtain the correct density of
photinos, the ratio $r \equiv m(R^0)/m_{\tilde{\gamma}}$ must fall between
about $ \sim 1.6 - 2$\cite{f:100}, which is in the range predicted on
the basis of the gluino and photino mass calcuations\cite{f}.  The
lifetime of the $R^0$ is then\cite{f} $ \tau_{R^0}
{\,\raisebox{-0.13cm}{$\stackrel{\textstyle>} {\textstyle\sim}$}\,}
(10^{-10} - 10^{-7}) \left(\frac{M_{sq}}{100 {\rm GeV}}\right)^4 $ sec
for $1.4 < M(R^0) < 2$ GeV.  This is comparable to the the $K_L^0 -
K_S^0$ 
lifetime range if $M_{sq} \sim 100$ GeV, or longer for heavier
squarks.  In ref. \cite{f:95} I discussed strategies for detecting or
excluding the existance of an $R^0$ with a lifetime so long it cannot be
detected by its decays.  Here I discuss several approaches appropriate
if the $R^0$ lifetime is in the $\sim 10^{-5} - 10^{-10}$s range.  

If $R^0$'s exist, beams for rare $K^0$ decay and $\epsilon'/\epsilon$
experiments would contain $R^0$'s.  The detectors designed to observe
$K^0$ decays can be used to study $R^0$ decays.  The $R^0$ production
cross section can be estimated in perturbative QCD when the $R^0$'s
are produced with $p_{\perp} {\,\raisebox{-0.13cm}{$\stackrel{\textstyle>}
{\textstyle\sim}$}\,} 1$ GeV.  However high-luminosity
beams are produced at low $p_{\perp}$ so pQCD cannot be used to
determine the $R^0$ flux in the beam.  The most important outstanding
phenomenological problem in studying light gluinos is to develop
reliable methods for estimating the $R^0$ production cross section in
the low $p_{\perp}$ region; this problem will be left for the future.
In the remainder of this paper I simply paramterize the ratio of $R^0$
to $K_L^0$ fluxes in a given beam at the production point by $p\cdot
10^{-4}$. 

The momentum in the $R^0$ rest frame of a hadron $h$, produced in the
two body decay $R^0 \rightarrow \tilde{\gamma} +~ h$, is $ P_h =
\sqrt{m_R^4 + m_{\tilde{\gamma}}^4 + m_h^4 - 2 m_R^2 m_{\tilde{\gamma}}^2 - 2
m_{\tilde{\gamma}}^2 m_h^2 - 2 m_h^2 m_R^2}/(2 m_R)$.  For the typical case
$1.6 < r ~{\,\raisebox{-0.13cm}{$\stackrel{\textstyle<}
{\textstyle\sim}$}\,} ~2$ and $m_{R^0} = 1.7$ GeV, $P_{\pi}\sim 500-600$ 
MeV.  This illustrates that, unless the $R^0$ is in the extreme high
end of its mass range and the photino is in the low end of its
estimated mass range, multihadron final states will be significantly
suppressed by phase space.  

While dominant with respect to phase space, two body decays are
suppressed by the approximate $C$-invariance of SUSY QCD.  The $R^0$
and $\tilde{\gamma}$ have $C=+1$ and $C=-1$
respectively\footnote{Familiar fermions are not eigenstates of $C$
because they have some non-vanishing conserved quantum number such as
charge or lepton number.  This is not true of the $R^0$ and photino.
Supersymmetry generators commute with the charge conjugation operator,
so the $\tilde{\gamma}$ and $R^0$ have the same $C$ as their
superpartners: the photon and $0^{++}$ glueball and $0^{-+}$ 
``$\eta_{\tilde{g}}$''.  Because SUSY and $C$ are broken, the mass
eigenstates in fact contain a small admixture of states with opposite
$C$.}, so that the $R^0$ can decay to a photino plus a single $C=+1$
meson such as a $\pi^0$ or $\eta$ only if charge conjugation is
violated.  In general $C$ and $P$ are violated, e.g., because the
superpartners of left and right chiral quarks are not mass degenerate.
The decay matrix element for $R^0 \rightarrow \tilde{\gamma} + 0^{-+}$
is proportional to $\frac{m(S_{uL})^2 - m(S_{uR})^2}{m(S_{uL})^2 +
m(S_{uR})^2}$ (and similar contributions from the $d$- and $s$- 
squarks, weighted with their charges and projected onto the flavor of
the pseudoscalar meson current).  Since the squark $L-R$
mass-splittings are a model-dependent aspect of SUSY-breaking,
we henceforth take the branching fraction of $R^0$ into two (three)
body final states to be a free parameter, $b_2$ ($b_3$).  The
$C$-allowed decays such as $R^0 \rightarrow \tilde{\gamma} \rho^0$ are
treated as three-body decays.  Since multibody decays are suppressed
by phase space, $b_2 + b_3  \approx 1$;  therefore bounding both $b_2$
and $b_3$ can rule out $R^0$'s.  

The most important three-body decay mode is $R^0 \rightarrow
\pi^+ \pi^- \tilde{\gamma}$.  Since the $R^0$ is a flavor singlet and the
$\tilde{\gamma}$ has photon-like couplings, the $\pi^+ \pi^- : \pi^0 \pi^0$
branching fractions are in the ratio 9:1.  Because of phase space
suppression, decays involving $K$'s and $\eta$'s can be neglected
compared to the  $\pi \pi \tilde{\gamma}$ final state.  Thus $R^0 \rightarrow
\pi^+ \pi^- \tilde{\gamma}$ accounts for $\sim 90\%$ of three-body
decays.   One can require $M(\pi^+ \pi^-) > M_K$ to reduce background
without a severe loss of signal: e.g., for $M_{R^0} = 1.7$ GeV and
$r=2$, 72\% of the $R^0 \rightarrow \pi^+ \pi^- \tilde{\gamma}$ decays would
pass this cut.  The branching fraction for decays meeting this cut is
therefore $0.65 ~b_3$.  

The dominant two-body decay channel is $R^0 \rightarrow \pi^0
\tilde{\gamma}$.  Searching for this decay is much like searching for the
decay $K^0_L \rightarrow \pi^0 \nu \bar{\nu}$.   Fortunately, the two
final states are readily distinguishable because a typical $\pi^0$
from $R^0$ decay has much larger $p_\perp$ than one from $K_L^0
\rightarrow \pi^0 \nu \bar{\nu}$, for which $p_{\perp}^{max} = 231$ MeV.
Furthermore, the $p_{\perp}$ spectrum of the pion in a two body
decay exhibits the striking Jacobean peak at $p_{\perp} = P_{\pi}$.
The existing limit on $br(K_L^0 \rightarrow \pi^0 \nu \bar{\nu})$ will
be used below to obtain some weak constraints on the $R^0$ lifetime
and production cross section.  Future experiments with a good
acceptance in the large $p_{\perp}$ region can place a much better
limit. 

Another interesting two-body decay is $R^0 \rightarrow \eta \tilde{\gamma}$.
Since $m(\eta) = 547~ {\rm MeV} > m(K^0) = 498$ MeV, there would be very
little background mimicking $\eta$'s in a high-resolution, precision
$K^0$-decay experiment.  Detecting $\eta$'s in the decay region of
one of these experiments, e.g., via their $\pi^+ \pi^- \pi^0$ or
$\pi^0 \pi^0 \pi^0$  final states whose branching fraction are 0.23
and 0.32, would be strong circumstantial evidence for an $R^0$.  The
relative strength of the $R^0 \rightarrow \pi^0 \tilde{\gamma}$ and $R^0
\rightarrow \eta \tilde{\gamma}$ matrix elements is determined by squark
masses and the $\eta \eta'$ mixing angle.  With the prefered
mixing angle and equal-mass $u$ and $d$ squarks the branching
ratio would be 0.23, if phase space suppression for the $\eta$ final
state could be neglected.  However since two body phase space $\sim
P_h$, in the $r$ region of interest the $R^0 \rightarrow \tilde{\gamma}
\eta$ decay is suppressed kinematically compared to $R^0 \rightarrow
\tilde{\gamma} \pi^0$. For $r = 1.6~(2.0)$ and $M_{R^0} = 1.7$ GeV, the
branching fraction for $R^0 \rightarrow \tilde{\gamma} \eta$ is reduced to
about $0.12~ b_2~ (0.17~  b_2)$ and drops rapidly for smaller
$M_{R^0}$.    

Although the rate for $R^0 \rightarrow (\eta \rightarrow \pi^+ \pi^-
\pi^0)~ \tilde{\gamma}$ may be only a few percent that of $R^0 \rightarrow \pi^0
\tilde{\gamma}$, both final states are comparably accessible because
experiments to study the single $\pi^0$ require a Dalitz conversion to
reduce background.  With full $p_{\perp}$ acceptance, $m_{\tilde{\gamma}}$
and $M_{R^0}$ can be determined with only and a handful of events in
both channels even though the momenta of the $R^0$ and $\tilde{\gamma}$ are
unknown.  The $p_{\perp}$ spectrum of a two body decay is strongly
peaked at $p_{\perp}^{max} = P_h$.  Thus determining $P_{\pi}$ and
$P_{\eta}$, gives two conditions fixing the two unknowns, $m(R^0)$ and
$m_{\tilde{\gamma}}$.  Determination of the ratio $m(R^0)/m_{\tilde{\gamma}}$ is
important to confirm or refute the proposal\cite{f:100} that relic
photinos are responsible for the bulk of the missing matter of the
Universe.    

\begin{figure}
\epsfxsize=\hsize
\epsffile{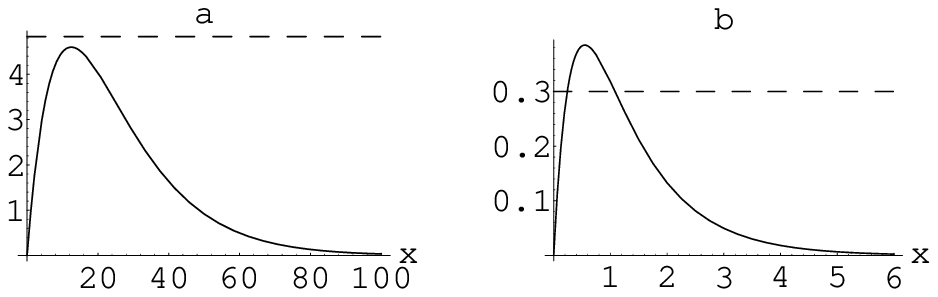}
\caption{Sensitivity function of (a) a typical $K_L^0$ beam and (b) an
NA48-like $K_S^0$ beam, with ${\cal S}^{lim} = 0.26$ indicated.}  
\label{kbeams}
\end{figure}

We can estimate the sensitivity of neutral kaon experiments to $R^0$'s
as follows.  The number of decays of a particle with decay length
$\lambda \equiv <\gamma \beta c \tau>$, in a fiducial region extending
from $L$ to $L+l$, is 
\eq
N = N_0 \left( e^{-\frac{L}{\lambda}} - e^{-\frac{L+l}{\lambda}}
\right),
\label{decays}
\en 
where $N_0$ is the total number of particles leaving the production
point.  In typical $K_L^0$ experiments\footnote{E.g., Fermilab's E799
and the $\frac{\epsilon'}{\epsilon}$ experiments KTeV and NA48 which
are scheduled to begin running during 1996 at FNAL and CERN.}, $L\sim
120$ m, $l \sim 12-30$ m, and $L/\lambda_{K_L^0} \sim 0.08 $, so $
e^{-\frac{L}{\lambda}} - e^{-\frac{(L+l)}{\lambda}} \approx
\frac{l}{\lambda}e^{-\frac{L}{\lambda}}$.  Denote the number of
reconstructed $R^0 \rightarrow \tilde{\gamma} X$ events by $N^R_X$ and 
denote the number of reconstructed $K_L \rightarrow Y$ events by
$N^K_Y$.  Then defining $br(R^0 \rightarrow \tilde{\gamma} X) \equiv b^R_X~
10^{-2}$ and $br(K_L \rightarrow Y) = b^K_Y 10^{-4}$, and idealizing
the particles as having a narrow energy spread, eq. (\ref{decays})
leads to: 
\eq
N^R_X \approx N^K_Y (~p~10^{-4})
\left(\frac{b^R_X~10^{-2}}{b^X_Y~10^{-4}}\right)
\left(\frac{\epsilon_X}{\epsilon_Y} \right)
\frac{<\gamma \beta \tau>_{K^0_L}}{<\gamma \beta \tau>_{R^0}}
exp[ -L/<\gamma \beta c \tau>_{R^0}],
\label{etas}
\en
where $\epsilon_X$ and $\epsilon_Y$ are the efficiencies for
reconstructing the final state particles $X$ and $Y$, $\gamma =
\frac{E}{m}$ is the relativistic time dilation factor, and $\beta = 
\frac{P}{E}$ will be taken to be 1 below.  Letting $x \equiv \frac{
\lambda_K}{\lambda_{R^0}} =\frac{<E_{K_L^0}>m_{R^0}
\tau_{K_L^0}}{<E_{R^0}>m_{K^0_L}\tau_{R^0}}$, and introducing the
``sensitivity function'' ${\cal S}(x) \equiv x~
exp[-L x/\lambda_{K^0_L}]$, eqn (\ref{etas}) implies that an
experiment with  
\eq
{\cal S}^{lim} \equiv \frac{100 ~b^K_Y}{p ~b^R_X} \frac{N^R_X}{N^K_Y} 
\frac{\epsilon_Y}{\epsilon_X}
\label{Slim}
\en
will restrict $x$ to be such that ${\cal S}(x) \le {\cal S}^{lim}$.
Thus the sensitivity of various experiments with the same
$L/\lambda_{K^0_L}$ can be directly compared by comparing their $
{\cal S}^{lim}$ values.  Fig. \ref{kbeams} shows ${\cal S}(x)$ for
$L/\lambda_{K_L^0} \sim 0.08 $.  The qualitative features are as
expected: an experiment with a large $K_L$ flux ($\frac{N_Y^K}{b_Y^K
\epsilon_Y}$) has a low ${\cal S}^{lim}$ and thus is sensitive to a
large range of $x \approx 4 \frac{\tau_{K_L^0}}{\tau_{R^0}} $. For
shorter lifetimes (large $x$), the $R^0$'s decay before reaching the
fiducial region, while for longer lifetimes (small $x$) the
probability of decay in the fiducial volume is too low for enough 
events to be seen. 

Consider first the Fermilab E799 experiment, which
obtained\cite{ktev:pi0nunubar} a 90\% cl  limit $br(K_L^0 \rightarrow
\pi^0 \nu \bar{\nu}) {\,\raisebox{-0.13cm}{$\stackrel{\textstyle<}
{\textstyle\sim}$}\,} ~5.8 ~10^{-5}$. In this case the $R^0$ final
state $X$ and the $K_L$ final state $Y$ both consist of a single
$\pi^0$ and missing energy.  Therefore $\frac{\epsilon_Y}{\epsilon_X }
$ is just the ratio of probabilities (which we will denote
respectively $f_K$ and $f_R$) for the $\pi^0$ to have $P_t$ in the
allowed range, $160 < P_t < 231$ GeV, in the two cases.  Taking
$br(R^0 \rightarrow \tilde{\gamma} \pi^0) \approx b_2$ and $br(K_L^0
\rightarrow \pi^0 \nu \bar{\nu}) \le ~5.8
~10^{-5}$\cite{ktev:pi0nunubar} means $b^R_Y = b_2~10^2$ and $b^K_Y < 
0.58$, so that we have
\eq
{\cal S}^{lim}_{E799} = \frac{0.58 f_K}{p~ b_2~ f_R }.
\label{xE799}
\en
With the spectrum $\frac{d\Gamma}{d E_{\pi^0}}$ used in ref. 
\cite{ktev:pi0nunubar}, $f_K = 0.5$.  For $R^0 \rightarrow \pi^0
\tilde{\gamma}$, $f_R = \frac{\sqrt{1 - (160)^2} - \sqrt{1 -
(231)^2}}{P_{\pi}} \approx (0.02 - 0.03)$, when $M_{R^0} = 1.4 - 2$ GeV
and $r$ is in the range $2.2 - 1.6$.  Taking $f_R = 0.025$ gives
$ {\cal S}^{lim}_{E799} =  11.6/(p b_2)$.  The peak of the function on
the lhs of eq. (\ref{xE799}) (see Fig. \ref{kbeams}a) occurs for $x = 
\frac{L}{\lambda_{K^0_L}}$, which is $\approx 12.5$.  Using $x \approx
4 \frac{\tau_{K_L^0}}{\tau_{R^0}} $, the peak sensitivity is for an
$R^0$ lifetime of $2~ 10^{-8}$s, for which the existing experimental
bound on $K_L^0 \rightarrow \pi^0 \nu \bar{\nu}$ yields a limit $p b_2 \le
2.4$.  Whether or not this is a significant restriction on $R^0$'s
can only be decided when reliable predictions for (or at least reliable
lower limits on) $b_2$ and the $R^0$ production cross section are in hand.

The next generation of $K_L^0$ experiments, KTeV and NA48, expect to
collect $\sim N^K_Y = 5~10^6$ reconstructed $K_L \rightarrow \pi^0 \pi^0$
events.  What sensitivity does this allow in searching for $R^0
\rightarrow \eta \tilde{\gamma}$, reconstructing the $\eta$ from its
$\pi^+ \pi^- \pi^0$ decay?  With a $\sim 5$ MeV resolution in the $\pi^+
\pi^- \pi^0$ invariant mass and negligible background between the
$K^0$ and $\eta$, three reconstructed $\eta$'s would be sufficient to
be convincing, so let us take $N^R_X = 3$.  We know $br(K_L^0
\rightarrow \pi^0 \pi^0) = 9~10^{-4}$ and $br(\eta\rightarrow
\pi^+ \pi^- \pi^0) = 0.23$, and take $br(R^0 \rightarrow \eta
\tilde{\gamma}) \approx 0.1~ b_2 $, so we have $b^K_Y = 9$ and $b^R_X \sim
2.3 ~b_2 10^{-2}$.  Thus ${\cal S}^{lim} = 2~10^{-2}
\frac{\epsilon_Y}{p b_2 \epsilon_X}$ where $\epsilon_Y$ is the
efficiency of reconstructing the ${\pi^0 \pi^0}$ final state of a
$K_L^0$ decay and $\epsilon_X$ is the efficiency for reconstructing
the $\pi^+ \pi^- \pi^0$ final state of an $\eta$.  $\epsilon_X$ needs
to be determined by Monte Carlo simulation.  If $p b_2 \sim 1$ and
$\epsilon_X$ is good enough that, say, ${\cal S}^{lim} = 3~10^{-2}/(p b_2)$,
such a sensitivity allows the range $0.03 < x < 102$ to be probed.  This
corresponds to an ability to discover $R^0$'s with a lifetime in the
range $\sim 2~ 10^{-9}- 0.7~ 10^{-5}$ sec.  Note that
in a rare $K_L^0$-decay experiment the flux of $K_L^0$'s is much
greater than for the $\frac{\epsilon'}{\epsilon}$ experiments, so
other things being equal a greater sensitivity can be achieved for a
comparable acceptance.  Unfortunately, E799 rejected the $\eta
\tilde{\gamma}$ final state. 

Use of an intense $K_S^0$ beam would allow shorter lifetimes to be
probed. The FNAL E621 experiment designed to search for the CP
violating $K_S^0 \rightarrow \pi^+ \pi^- \pi^0$ decay had a high
$K_S^0$ flux and a decay region close to the production
target.  However its 20 MeV invariant mass resolution may be
insufficient to adequately distinguish $\eta$'s from $K^0$'s.
To estimate the sensitivity of, e.g., the NA48 detector we must return 
to eq. (\ref{decays}), since for the planned $K_S^0$ beam
$\lambda_{K_S^0} \approx L \approx l/2$.  In this case $x_S \equiv
\frac{<E_{K_S^0}>m_{R^0}\tau_{K_S^0}}{<E_{R^0}>m_{K^0_S}\tau_{R^0}}
\approx \frac{4~\tau_{K_S^0}}{\tau_{R^0}}$, must
satisfy  
\begin{eqnarray}
{\cal S}^S(x) & = & \left( e^{-\frac{Lx_S}{\lambda_{K_S^0}}} -
e^{-\frac{(L+l)x_S}{\lambda_{K_S^0}}} \right) < \nonumber \\
 {\cal S}_{lim}^S & \equiv & \left( e^{-\frac{L}{\lambda_{K_S^0}}} -
e^{-\frac{(L+l)}{\lambda_{K_S^0}}} \right)\frac{br(K_S^0 \rightarrow
\pi^0 \pi^0)}{b^R_X  ~10^{-2}~ p~10^{-4}}\frac{N_{R^{+-0}}}{N_{K_S^{00}}}
\frac{\epsilon^{00}}{\epsilon^{+-0}}.
\end{eqnarray}
Taking the same production rate and efficiencies as before, and
assuming $\sim 10^7$ reconstructed $K_S^0 \rightarrow \pi^0 \pi^0$
decays, gives ${\cal S}^{lim}_S =  0.26$.  ${\cal S}_S(x)$ is shown in
Fig. \ref{kbeams}b.  The sensitivity range is $0.19 < x_S < 1.3$ for $pb_2
= 1$;  this corresponds to the lifetime range $ 3~10^{-10} - 2~10^{-9}$s. 

Thus for $p b_2 \approx 1$ the next generation of $\epsilon'/\epsilon$
experiments will be able to see $R^0$'s in the lifetime range $3~
10^{-10} - 0.7 ~10^{-5}$ sec.  The greatest sensitivity is for
$\tau_{R^0} = 2~10^{-8}$ sec; for this lifetime, values of $pb_2$ as
small as $\sim 6~ 10^{-3}$ should be accessible.  For a given $p$,
even better sensitivity is possible, using the final state $\pi^+
\pi^- \tilde{\gamma}$ with $m(\pi^+ \pi^-)>m_K$, if $b_3 \ge b_2/8$.
If we assume the background to this mode is low enough that observing
$\sim 10$ events with $m(\pi^+ \pi^-)>m_K$ is sufficient for
detection, the factor $\frac{N_X^R}{b_X^R \epsilon_X}$ appearing in
eq. (\ref{Slim}) is reduced by the factor $\frac{10/3}{(0.65
b_3)/(0.023 b_2)}$.  Thus ${\cal S}^{lim}$ is reduced by the factor
$0.12 b_2/b_3$ compared to the $R^0 \rightarrow \eta \tilde{\gamma}$
search.  Hence, unless $p << 1$, the planned $\epsilon'/\epsilon$
experiments will be sensitive to nearly the entire lifetime range of
interest below $\sim 10^{-5}$ sec independently of the relative
importance of 2- and 3-body decays of the $R^0$.

Turning now to other $R$-hadrons, the ground-state $R$-baryon is the
flavor singlet scalar $uds \tilde{g}$ bound state denoted $S^0$.  On
account of the very strong hyperfine attraction among the quarks in
the flavor-singlet channel\cite{f:52}, its mass is about $210 \pm 20$
MeV lower than that of the lowest R-nucleons.  The mass of the $S^0$
is almost surely less than $m(\Lambda) + m(R^0)$, so it cannot decay
through strong interactions.  As long as $m(S^0)$ is less than $m(p) +
m(R^0)$, the $S^0$ must decay to a photino rather than $R^0$ and
would have an extremely long lifetime since its decay requires a
flavor-changing-neutral-weak transition. The $S^0$ could even be
stable, if $m(S^0) - m(p) - m(e^-) < m_{\tilde{\gamma}}$ and $R$-parity is a
good quantum number\footnote{If the baryon  resonance known as the
$\Lambda(1405)$ is a ``cryptoexotic'' flavor singlet bound state of
$udsg$, one would expect the corresponding state with gluon replaced
by a light gluino to be similar in mass.  In this case the $S^0$ mass
would be $\sim 1 \frac{1}{2}$ GeV and the $S^0$ would be stable as
long as the photino is heavier than $\sim 600$ MeV, as it would be
expected to be if photinos account for the relic dark matter.}.  This
is not experimentally excluded\cite{f:51,f:95} because the $S^0$
probably does not bind to nuclei.  The two-pion-exchange force, which
is attractive between nucleons, is repulsive between $S^0$ and nucleons
because the mass of the intermediate $R_{\Lambda}$ or $R_{\Sigma}$ is
much larger than that of the $S^0$.  

If the $S^0$ is stable, it provides a possible explanation for the
several very high energy cosmic ray events which have been recently
observed\cite{akeno_flyseye}.  Greisen-Zatsepin-Kuzmin (GZK) pointed
out\cite{GZK} that the cross section for proton
scattering from the cosmic microwave background radiation is very
large for energies above $\sim 10^{20}$ eV, because at such energies
the $\Delta(1230)$ resonance is excited.  If cosmic 
ray protons are observed with larger energies than the GZK bound they
must have originated within about 30 Mpc of our galaxy.
Since there are no good candidates for ultra-high energy cosmic
ray sources that close, the observed events with $E \sim 3~10^{20}$
eV\cite{akeno_flyseye} have produced a puzzle for astrophysics.
However the threshold for producing a resonance of mass 
$M^*$ in $\gamma({\rm 3^o K}) + S^0$ collisions is a factor
$\frac{m_{S^0}}{m_p} \frac{( M^* - M_{S^0})}{(1230-940) {\rm MeV}}$
larger than the threshold in $\gamma({\rm 3^o K}) + p$ collisions.
Taking $m(R^0) = 1.7$ GeV, $m_{\tilde{\gamma}}$ must lie in the range $0.8
\sim 1.1$ GeV to account for the relic dark matter.  If $m_{S^0}
\approx m_p + m_{\tilde{\gamma}}$ we have $m(S^0) \sim 1.8 - 2.1$ GeV.
Since the photon couples as a flavor octet, the resonances excited in
$S^0 \gamma$ collisions are flavor octets.  Since the $S^0$ has
spin-0, only a spin-1 $R_{\Lambda}$ or $R_{\Sigma}$ can be produced
without an angular momentum barrier.  There are two $R$-baryon flavor
octets with $J=1$, one with total quark spin 3/2 and the other with
total quark spin 1/2, like the $S^0$.  Neglecting the mixing between
these states which is small, their masses are about 385-460 and
815-890 MeV heavier than the $S^0$, respectively\cite{f:52}.  Thus the
GZK bound is increased by a factor of 2.4 - 6.5, depending on 
which $R$-hyperons are strongly coupled to the $\gamma S^0$ system. 
Therefore, if $S^0$'s are stable they naturally increase the GZK bound
enough to be compatible with the extremely high energy cosmic rays
reported in \cite{akeno_flyseye} and references therein.  

The $S^0$ can be produced via a reaction such as $ K ~ p \rightarrow
R^0 ~ S^0 + X$, or can be produced via decay of a higher mass
$R$-baryon such as an $R$-proton produced in $ p ~ p \rightarrow R_p ~
R_p + X$.  In an intense proton beam at relatively low energy, the
latter reaction is likely to be the most efficient mechanism for
producing $S^0$'s, as it minimizes the production of ``extra'' mass.
One strategy for finding evidence for the $S^0$ would be to perform an
experiment like that of Gustafson et al\cite{gustafson}, in which a
neutral particle's velocity is measured by time of flight and its
kinetic energy is measured in a calorimeter. This allows its mass to
be determined via the relation $KE = m(\frac{1}{\sqrt{1- \beta^2}}
-1)$.  On account of limitations in time of flight resolution and
kinetic energy measurement, ref. \cite{gustafson} was only able to
study masses $> 2$ GeV, below which the background from neutrons
became too large.  An interesting aspect of using a primary proton
beam at the Brookhaven AGS, where the available cm energy is limited
($p_{beam} \sim 20$ GeV), is that pair production of $S^0$'s probably
dominates associated production of $S^0$-$R^0$ and production of $R^0$
pairs, due to the efficiency from an energy standpoint of packaging
baryon number and R-parity together in an $S^0$ or $R_p$.  The
expectation that $S^0$'s are produced in pairs gives an extra
constraint which can help discriminate against the neutron background
in such a search.  It is also helpful that, for low energy $S^0$'s, the
calorimetric determination of the $S^0$ kinetic energy is not smeared
by conversion to $R^0$ because of the $t_{min}$ required for a
reaction like $S^0 ~ N \rightarrow R^0 +\Lambda + N' + X$.  Although
the $S^0$ has approximately neutron-like interaction with matter, its
cross section could easily differ from that of a neutron by a factor
of two or more, so that systematic effects on the calorimetry of the
unknown $S^0$ cross section should be considered. 

If the $R^0$ is too long-lived to be found via anomalous decays
in kaon beams and the $S^0$ cannot be discriminated from a neutron,
a dedicated experiment studying two-body reactions of the type
$R^0 + N \rightarrow K^{+,0}+ S^0$ could be done.  Depending on the
distance from the primary target and the nature of the detector, the
backgrounds would be processes such as $K^0_L + N \rightarrow K^{+,0} +
n$, etc.  If the final state neutral baryon is required to rescatter,
and the momentum of the kaon is determined, and time of flight is used
to determine $\beta$ for the incident particle, all with sufficient
accuracy, one would have enough constraints to establish that one was
dealing with a two-body scattering and to determine the $S^0$ and
$R^0$ masses.  Measuring the final neutral baryon's kinetic energy
would give an over-constrained fit which would be helpful.

Light $R$-hadrons other than the $R^0$ and $S^0$ will decay, most via
the strong interactions, into one of these.  However since the
lightest $R$-nucleons are only about $210 \pm 20$ MeV heavier than the
$S^0$, they would decay weakly, mainly to $S^0 \pi$.  The $R$-nucleon
lifetimes should be of order $2~ 10^{-11} - 2~ 10^{-10}$ sec, by
scaling the rates for the analog weak decays $\Sigma^- \rightarrow
n~\pi^-,~\Lambda^- \rightarrow p ~\pi^-$ and $\Xi^- \rightarrow
\Lambda ~\pi^-$ by phase space.  Existing experimental
limits\cite{f:95} do not apply to the lifetime region and kinematics
of interest.  Silicon microstrip detectors developed for charm studies
are optimized for the lifetime range $(0.2 - 1.0)~10^{-12}$ sec. 
Moreover unlike ordinary hyperon decay, there is at most one charged
particle in the final state, except for very low branching fraction
reactions such as $R_n \rightarrow S^0 ~\pi^-~ e^+ ~\nu_e$, or $R_n
\rightarrow S^0~\pi^0$ followed by $\pi^0 \rightarrow \gamma~ e^+~
e^-$.  In order to distinguish the decay $R_p \rightarrow S^0 \pi^+$
from the much more abundant background such as $\Sigma^+ \rightarrow n
~ \pi^+$, which has a similar energy release, one could rescatter the
final neutral in order to get its direction.  Then with sufficiently
accurate knowledge of the momentum of the initial charged beam and the
momentum (and identity) of the final pion, one has enough constraints
to determine the masses of the initial and final baryons.  The
feasibility of such an experiment is worth investigating.  Even
without the ability to reconstruct the events, with sufficiently good
momentum resolution for the initial and final charged particles, one
could search for events which are not consistent with the kinematics
of known processes such as $\Sigma^+ \rightarrow \pi^+ n$, and then
see if they are consistent with the two body decay expected here.  One
other charged $R$-baryon could be strong-interaction stable, the
$R_{\Omega^-}$.  Assuming its mass is 940 MeV ($= m(\Omega^-)-m(N) +
210$ MeV) greater than the $S^0$ mass, it decays weakly to $R_{\Xi} +
\pi$ or $R_{\Sigma}+ K$, with the $R_{\Xi}$ or $R_{\Sigma}$ decaying
strongly to $S^0 K$ or $S^0 \pi$ respectively.  This would produce a
more distinctive signature than the $R$-nucleon decays, but at the
expense of the lower production cross section.

In addition to the new hadrons expected when there are light gluinos
in the theory, there are many other consequences of light gluinos.
Since gluinos in this scenario live long enough that they hadronize
before decaying to a photino, they produce jets similar to those
produced by the other light, colored quanta: gluons and quarks.  In
$Z^0$ decay, only 4- and more- jet events are modified and the
magnitude of the expected change is smaller than the uncertainty in
the theoretical prediction\cite{f:82}.  Calculation of the 1-loop
corrections to the 4-jet amplitudes is needed.  In $p \bar{p}$
collisions, there is a difference between QCD with and without gluinos
already in 1-jet cross sections.  However absolute predictions are
more difficult than for $Z^0$ decay since they rely on structure
functions which have so far been determined assuming QCD without
gluinos.   Less model dependent might be to search for differences in
the expected {\it relative} $n$-jets cross sections\cite{f:82}.  Other
indirect consequences of light gluinos are not presently capable of
settling the question as to whether light gluinos exist, since they
all rely on detailed understanding of non-perturbative aspects of QCD.
Length restrictions prevent reviewing them here.

{\bf Acknowledgements:}  I am indebted to many people for information
and helpful discussions and suggestions including M. Calvetti, J. Conway, 
T. Devlin, J. Kane, S. Lammel, L. Littenberg, I. Mannelli, J. Rosner, M.
Schwartz, S. Somalwar, G. Thomson, Y. Wah, W. Willis, B. Winstein,
and M. Witherell.   




\end{document}